\documentclass[twocolumn,twocolappendix]{aastex701}

	\usepackage{amsmath}
	\usepackage{amssymb}
	\usepackage{bm}
    \usepackage{listings}
    \usepackage{enumitem}
	
	\let\vec\bm

	\newcommand{\diff}{\ensuremath{\mathrm{d}}}
	\newcommand{\e}{\mathrm{e}}
	\newcommand{\I}{\mathrm{i}}

	\newcommand{\p}{\prime}

	\newcommand{\rhos}{\rho_\mathrm{s}}
	\newcommand{\rs}{r_\mathrm{s}}

	\newcommand{\vc}{v_\mathrm{circ}}
	
	\newcommand{\Msol}{\mathrm{M}_\odot}
	\newcommand{\kpc}{\mathrm{kpc}}
	\newcommand{\pc}{\mathrm{pc}}
	\newcommand{\kms}{\mathrm{km}\,\mathrm{s}^{-1}}

	\graphicspath{{./}{figures/}}
		
\begin{document}
			
			\title{Testing warm dark matter with kinematics of the smallest galaxies}

			\author[orcid=0000-0003-3808-5321,sname='Delos']{M. Sten Delos}
			\affiliation{Carnegie Observatories, 813 Santa Barbara Street, Pasadena, CA 91101, USA}
			\email[show]{mdelos@carnegiescience.edu}  
			
			\author[orcid=0009-0002-1233-2013,sname='Ahvazi']{Niusha Ahvazi}
			\affiliation{Department of Astronomy, University of Virginia, 530 McCormick Road, Charlottesville, VA 22904, USA}
			\email{nahvazi@virginia.edu}  
			
			\author[orcid=0000-0001-5501-6008,sname='Benson']{Andrew Benson}
			\affiliation{Carnegie Observatories, 813 Santa Barbara Street, Pasadena, CA 91101, USA}
			\email{abenson@carnegiescience.edu}  
			
			\begin{abstract}
				
				Every dark matter halo forms with a $\rho\propto r^{-1.5}$ density cusp at its center. For warm dark matter (WDM), these \textit{prompt cusps} can be massive enough to influence the kinematics of dwarf galaxies. By implementing prompt cusps in the \textsc{Galacticus} galaxy formation model, we show that the measured velocity dispersions of Tucana V and Triangulum II are serious outliers for dwarf galaxies arising in WDM models. For thermal-relic dark matter, the three faintest Milky Way satellites together constrain the particle mass to be $m_\chi>5.8$~keV at 95 percent confidence or $m_\chi>9.4$~keV at 90 percent confidence. Improved velocity dispersion measurements for these systems could greatly refine this constraint, as could identification and kinematic characterization of more such galaxies.
				
			\end{abstract}
			
		
		
		\section{Introduction}\label{sec:intro}
		
		Dark matter comprises most of the matter in the universe, but its particle nature remains unknown. The primordial velocity distribution of the dark matter could provide an important clue toward identifying it. The dark matter production mechanism could have imparted a substantial velocity dispersion. If the dark matter was in thermal contact with the Standard Model in the past, it could also have acquired significant thermal motion. In these warm dark matter (WDM) scenarios, the random velocities would gradually erase density variations as particles are able to stream freely out of overdense or underdense regions. Thermal motions effectively smooth the density field of the early universe on some characteristic free-streaming length scale.
		
		Observational evidence for small-scale initial density perturbations is often used to place limits on the free-streaming scale and hence the dark matter velocity distribution. If the dark matter is a thermal relic (initially in equilibrium with the Standard Model), then its velocity distribution is linked to the particle mass $m_\chi$. In this case, a range of prior works have placed lower limits around $m_\chi\gtrsim 6$~keV based on the abundance of Milky Way satellite galaxies \citep{2025ApJ...986..127N}, perturbations induced by low-mass halos to strong gravitational lenses \citep{2024MNRAS.535.1652K,2025arXiv251107513G} and stellar streams \citep{2021JCAP...10..043B}, and density variations inferred from the Lyman $\alpha$ forest \citep{2023PhRvD.108b3502V,2024PhRvD.109d3511I}.
		
		However, the free-streaming scale is also associated with the formation of prompt cusps \citep{2019PhRvD.100b3523D,2023MNRAS.518.3509D,2023JCAP...10..008D,2024MNRAS.52710802O}. These $\rho\propto r^{-1.5}$ density cusps arise at that scale from the collapse of smooth peaks in the initial density field. They are the first and smallest elements of cosmic structure, around which all larger dark matter systems grow.
		In this article, we use the prompt cusps that would reside at the centers of Milky Way satellite galaxies to place complementary limits on the dark matter mass.
		
		\citet{2023MNRAS.522L..78D} first explored using prompt cusps as a probe of WDM. Recently, \citet{2025ApJ...993...93D} used cosmological simulation results to characterize how the central prompt cusp of each dark matter halo is set by the free-streaming scale. We implement this characterization in the \textsc{Galacticus} semi-analytic galaxy model in order to evaluate the expected prompt cusps of Milky Way satellites. We consider the three smallest and faintest satellites -- Segue~1, Triangulum II (Tri~II), and Tucana V (Tuc~V) -- and compare the observed kinematics of these galaxies with those of their \textsc{Galacticus} counterparts selected by absolute magnitude, half-light radius, and orbital pericenter. Through this comparison, we constrain $m_\chi>5.8$~keV at 95 percent confidence. At 90 percent confidence, the constraint improves to $m_\chi>9.4$~keV.
		
\begin{figure}
			\centering
			\includegraphics[width=\columnwidth]{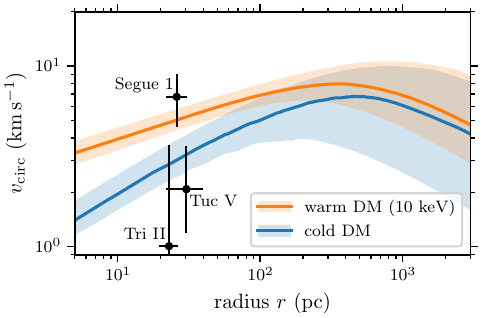}
			\caption{Illustration of the analysis in this work. In color, we show radial profiles of the circular orbit velocity for satellite galaxies similar to Segue~1, Tri~II, and Tuc~V produced with \textsc{Galacticus}. We show the median and 68 percent scatter at each radius, and different colors correspond to different dark matter (DM) models. For the same galaxies, the points with $1\sigma$ error bars mark the observationally inferred circular orbit velocity at the half-light radius.}
			\label{fig:vcirc}
		\end{figure}

		Figure~\ref{fig:vcirc} illustrates how this test operates. For WDM, the prompt cusp is associated with a high concentration of mass at the center of the dark matter halo. This mass would result in faster stellar orbits compared to a scenario with cold dark matter (CDM) of much higher $m_\chi$. The difference in circular orbit speeds between WDM (with a massive prompt cusp) and CDM (with a negligibly low-mass prompt cusp) is most pronounced at the smallest radii. Since galactic structures and kinematics are only observed in projection, the speed of a circular orbit is most robustly estimated near the half-light radius of a galaxy. This is why we focus on the spatially smallest galaxies. We also focus on faint galaxies in order to minimize the potential for star formation feedback to alter the initial dark matter distribution.
		
		This article is organized as follows.
        In section~\ref{sec:satellites}, we discuss our selection of Milky Way satellite galaxies and the observational inferences of these galaxies that we use.
        Section~\ref{sec:model} describes how we model theoretical analogues of these galaxies, while in section~\ref{sec:analogues} we explore the properties of these analogues.
        In section~\ref{sec:limits}, we compare the model to the observational inferences to derive constraints on dark matter.
        We conclude in section~\ref{sec:conclusion}.
        Three appendices follow: appendix~\ref{sec:simulations} analyzes cosmological simulations to inform our modeling and test some of the results; appendix~\ref{sec:distributions} further tests the properties of the theoretical galaxy analogues; and appendix~\ref{sec:varyA} examines how sensitive the dark matter constraints are to modeling error in the prompt cusp coefficients.

		\section{The smallest and faintest Milky Way satellites}\label{sec:satellites}

\begin{figure}
			\centering
			\includegraphics[width=\columnwidth]{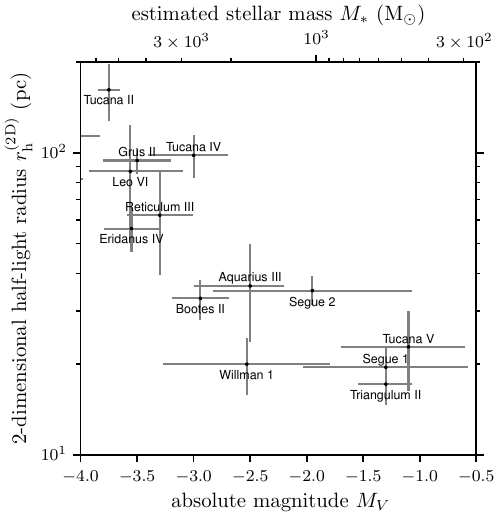}
			\caption{Low-luminosity Milky Way satellites drawn from the LVDB (v1.0.6), shown in terms of absolute $V$-band magnitude $M_V$ and azimuthally averaged 2-dimensional half-light radius $r_\mathrm{h}^\mathrm{(2D)}$ (with $1\sigma$ error bars). In this work we consider Segue~1, Tri~II, and Tuc~V, which are the faintest and smallest confirmed Milky Way satellite galaxies.}
			\label{fig:dwarfs}
		\end{figure}

        Figure~\ref{fig:dwarfs} shows the low-luminosity end of the Milky Way satellite galaxy distribution, drawn from the Local Volume Database \citep[LVDB; ][]{2025OJAp....8E.142P}.\footnote{\url{https://github.com/apace7/local_volume_database}; we use version 1.0.6 \citep{andrew_pace_2025_17291209} and restrict our consideration to systems for which \lstinline[basicstyle=\ttfamily]|host| is \lstinline[basicstyle=\ttfamily]|"mw"| and \lstinline[basicstyle=\ttfamily]|confirmed_galaxy| is \lstinline[basicstyle=\ttfamily]|1|.}
        Since the prompt cusp is most relevant at the smallest radii, we are interested in the least spatially extended galaxies. Moreover, to minimize the possibility of baryonic feedback affecting the dark matter distribution, we are interested in the faintest galaxies. Therefore, we select Segue~1, Tri~II, and Tuc~V.\footnote{Although Willman 1 is comparably small, it has a somewhat higher stellar mass, and it is listed by \citet{2019ARA&A..57..375S} as a galaxy ``for which published kinematics may not reliably translate to masses''.}
		
\begin{figure}
			\centering
			\includegraphics[width=\columnwidth]{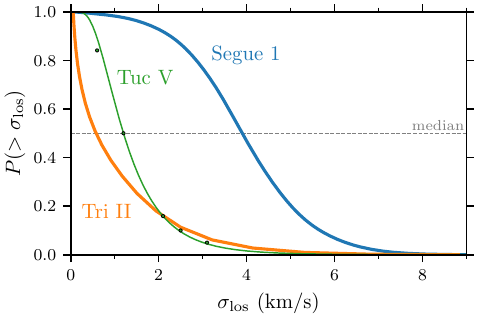}
			\caption{Cumulative posterior distributions of the line-of-sight velocity dispersion of the dwarf galaxies that we analyze. For Segue~1 and Tri~II, we take the distributions directly from \citet{2011ApJ...733...46S} and \citet{2022MNRAS.514.1706B}, respectively. For Tuc~V, the points correspond to the values reported by \citet{2024ApJ...968...21H}, and we adopt a closely matching log-normal distribution (green curve).}
			\label{fig:sigmapost}
		\end{figure}

        For our analysis, we focus on four main parameters of each galaxy:
        \begin{enumerate}[label={(\arabic*)},nosep]
            \item The absolute $V$-band magnitude $M_V$. $M_V$ is a proxy for the galaxy's stellar mass. Through galaxy formation physics, the stellar mass in turn provides approximate information about the dark matter halo mass \citep[e.g.][]{2020ApJ...893...48N,2022MNRAS.516.3944M,2023MNRAS.519..871Z}.
            \item The spherically averaged half-light radius $r_\mathrm{h}$, which we estimate as $r_\mathrm{h} = (4/3)r_\mathrm{h}^\mathrm{(2D)}$ \citep{2010MNRAS.406.1220W}, where $r_\mathrm{h}^\mathrm{(2D)}$ is the azimuthally averaged 2-dimensional half-light radius. $r_\mathrm{h}$ is important because it sets the radius at which the halo mass distribution can be most precisely inferred from kinematics.
            \item The orbital pericenter radius $r_\mathrm{p}$. The galaxy's orbit within the Milky Way halo is important because it sets how tidal forces reshape the halo density profile. $r_\mathrm{p}$ is the most important orbital parameter for this tidal evolution, and $r_\mathrm{p}$ also tends to be one of the most tightly constrained orbital parameters from observations.
            \item The circular orbit velocity at the half-light radius, $\vc(r_\mathrm{h})$, which we estimate as $\vc(r_\mathrm{h})=\sqrt{3}\sigma_\mathrm{los}$ \citep{2010MNRAS.406.1220W}, where $\sigma_\mathrm{los}$ is the line-of-sight velocity dispersion. $\vc(r_\mathrm{h})$ is the key property that is influenced by the presence or absence of a massive prompt cusp at the galaxy's center.
        \end{enumerate}
        For Segue~1, Tri~II, and Tuc~V, we make use of the following observational inferences of these parameters.
        \begin{itemize}[nosep]
            \item \textbf{Segue~1}: We take $M_V=-1.3\pm 0.73$ and $r_\mathrm{h}^\mathrm{(2D)}=19.5\substack{+3.4 \\ -3.1}$~pc from the LVDB (v1.0.6) \citep[based on][]{2007ApJ...654..897B,2018ApJ...860...66M} along with $r_\mathrm{p}=21\substack{+4 \\ -5}$~kpc from \citet{2021ApJ...916....8L}. For $\sigma_\mathrm{los}$, we use the posterior distribution from \citet[Figure 6 with binary correction]{2011ApJ...733...46S}.
            \item \textbf{Tri~II}: We take $M_V=-1.3\substack{+0.23 \\ -0.25}$ and $r_\mathrm{h}^\mathrm{(2D)}=17.2\substack{+2.6 \\ -2.5}$~pc from the LVDB (v1.0.6) \citep[based on][]{2017AJ....154..267C,2024ApJ...967...72R} along with $r_\mathrm{p}=12\pm 1$~kpc from \citet{2021ApJ...916....8L}. For $\sigma_\mathrm{los}$, we use the posterior distribution from \citet[Figure 10 with Raghavan prior]{2022MNRAS.514.1706B}.
            \item \textbf{Tuc~V}: We take $M_V=-1.1\substack{+0.5 \\ -0.6}$ and $r_\mathrm{h}^\mathrm{(2D)}=22.8\substack{+7.1 \\ -6.5}$~pc from the LVDB (v1.0.6) \citep[based on][]{2020ApJ...892..137S} along with $r_\mathrm{p}=38\pm 14$~kpc from \citet{2021ApJ...916....8L}. For $\sigma_\mathrm{los}$, we adopt a log-normal distribution centered at $1.2~\kms$ with $0.56$ e-fold standard deviation, which closely matches the constraints reported by \citet{2024ApJ...968...21H}.
        \end{itemize}
        Table~\ref{tab:data} summarizes the values of $M_V$, $r_\mathrm{h}^\mathrm{(2D)}$, and $r_\mathrm{p}$. For each of these parameters, we model the uncertainty with two half-normal distributions. For $\sigma_\mathrm{los}$, figure~\ref{fig:sigmapost} shows the posterior distributions.
        As noted above, we estimate $r_\mathrm{h} = (4/3)r_\mathrm{h}^\mathrm{(2D)}$ and $\vc(r_\mathrm{h})=\sqrt{3}\sigma_\mathrm{los}$.

        \begin{table}
            \caption{Observational inferences used in this work.}\label{tab:data}
            \begin{tabular*}{\columnwidth}{@{\extracolsep{\fill}} c c c c }
		          \hline
                Galaxy & $M_V$ & $r_\mathrm{h}^\mathrm{(2D)}$ (pc) & $r_\mathrm{p}$ (kpc)\\
		          \hline
                Segue~1 & $-1.30\substack{+0.73 \\ -0.73}$ & $19.5\substack{+3.4 \\ -3.1}$ & $21\substack{+4 \\ -5}$ \\
                Tri~II & $-1.30\substack{+0.23 \\ -0.25}$ & $17.2\substack{+2.6 \\ -2.5}$ & $12\substack{+1 \\ -1}$ \\
                Tuc~V & $-1.1\substack{+0.5 \\ -0.6}$ & $22.8\substack{+7.1 \\ -6.5}$ & $38\substack{+14 \\ -14}$ \\
            \end{tabular*}
            \tablecomments{$M_V$ is the absolute $V$-band magnitude, $r_\mathrm{h}^\mathrm{(2D)}$ is the azimuthally averaged 2-dimensional half-light radius, and $r_\mathrm{p}$ is the orbital pericenter radius.}
        \end{table}
        
        \section{Modeling prompt cusps and galaxies}\label{sec:model}

        We now discuss how we model the prompt cusps of these satellite galaxies.
        \citet{2025ApJ...993...93D} provided a prescription for calculating the central prompt cusp of each dark matter halo as a function of halo mass and time.
        However, this prescription is for field halos, and estimating the prompt cusp of a subhalo requires knowledge of the subhalo's mass at its initial infall time or earlier (when it was still a field halo).
        Therefore, our overall analysis strategy is to trace the past histories of galaxies similar to Segue~1, Tri~II, and Tuc~V.
        
        We use the \textsc{Galacticus} semi-analytic model \citep{2011ascl.soft08004B,2012NewA...17..175B}\footnote{\url{https://github.com/galacticusorg/galacticus}; the version that we use is at \url{https://github.com/galacticusorg/galacticus/tree/0925ee9aae2d06dbe86d1f4dee29e93087f62dcf}.} to simulate the full subhalo population of a Milky Way-like galaxy with a present-day mass of $10^{12}~\Msol$. \textsc{Galacticus} generates the tree of halo progenitors of the main halo using a prescription based on excursion set theory -- specifically it uses the modified merger rates of \cite{2008MNRAS.383..557P}, with the extended modifier function proposed by \cite{2025MNRAS.541.3713N} with parameters\footnote{The additional parameters $\gamma_4$ and $\gamma_5$ of the \cite{2025MNRAS.541.3713N} modifier were not used here.} $(G_0,\gamma_1,\gamma_2,\gamma_3)=(1.14,-0.33,0.059,0.65)$ found by constraining \textsc{Galacticus} merger rates to match progenitor mass functions of halos measured in the MDPL \citep{2012MNRAS.423.3018P} and Caterpillar \citep{2016ApJ...818...10G} simulations. Within this tree, each branching event corresponds to infall of a subhalo into another halo, and that subhalo's orbit is integrated within the gravitational potential of its host while accounting for dynamical friction and tidal evolution \citep{2014ApJ...792...24P,2020MNRAS.498.3902Y,2024PhRvD.110b3019D}.

        For computational efficiency, we only track progenitor halos down to a mass resolution threshold of $10^7~\Msol$ \citep[as suggested by][]{2024MNRAS.529.3387A}.\footnote{Note however that we track subhalos through tidal evolution down to arbitrarily low masses.} We have updated \textsc{Galacticus} to use the cusp-halo relation of \citet{2025ApJ...993...93D} to assign a prompt cusp to each halo at the time that it crosses the resolution threshold. This prescription provides the median cusp coefficient $A$, and we additionally include lognormally distributed scatter in $A$ following appendix~\ref{sec:simulations}. Since prompt cusps do not evolve significantly in time \citep{2023MNRAS.518.3509D}, the same $A$ remains associated with the halo up to the present day.

        Halo concentration parameters are evolved using the prescription of \citet{2021ApJ...908...33J} calibrated to match the concentration-mass relation of \citet{2016MNRAS.460.1214L}.
        The density profile of a halo is taken to follow the cusp-NFW form \citep{2025ApJ...993...93D},
        \begin{align}\label{cuspNFW}
            \rho(r)=\frac{\sqrt{y^2+x}}{x^{1.5}(1+x)^2}\rhos,
            \quad
            x\equiv \frac{r}{\rs}, \quad y \equiv \frac{A}{\rhos\rs^{1.5}},
        \end{align}
        which transitions from a $\rho=Ar^{-1.5}$ prompt cusp at small radii to a Navarro-Frenk-White (NFW) profile \citep{1996ApJ...462..563N,1997ApJ...490..493N} at large radii.
        Here $\rs$ and $\rhos$ are scale radius and density parameters that are set to match the halo mass and concentration. The concentration parameter is defined as $c=r_\mathrm{vir}/r_{-2}$, where $r_\mathrm{vir}$ is the virial radius and $r_{-2}$ is the radius at which $\diff\ln\rho/\diff\ln r=-2$; note that generally $\rs\geq r_{-2}$ (with equality only for $y=0$).
        For subhalos, the density profile is then modified by tidal forces according to the tidal heating prescription of \citet{2014ApJ...792...24P} calibrated to $N$-body simulations as in \citet{2020MNRAS.498.3902Y}.

        Finally, \textsc{Galacticus} employs a comprehensive and physically motivated prescription for galaxy formation within each halo. In this work we adopt the model of \citet{2024MNRAS.529.3387A}, which includes treatments for gas accretion and cooling, star formation, stellar feedback, and reionization. Gas cooling follows the \citet{White-Frenk1991} formalism, using metallicity-dependent atomic cooling rates from \textsc{cloudy} \citep[v23.01,][]{Gunasekera2023}, supplemented by molecular hydrogen (H$_2$) cooling computed using the \citet{Abel1997} chemical network and \citet{Galli1998} cooling functions. The circumgalactic medium is modeled as a uniform-density sphere at the halo virial temperature, tracking H$_2$ formation, photodissociation by the evolving ultraviolet background radiation of \citet{FaucherGiguere2020}, and self-shielding following \citet{Safranek-Shrader2012}. These extended cooling physics, combined with a reionization model based on \citet{Benson2020} in which the intergalactic medium is instantaneously heated at $z\simeq10$ and subsequently cools, captures the suppression of galaxy formation in the lowest-mass halos. Gas accretion follows the filtering-mass prescription of \citet{Naoz-Barkana2007}, while star formation in the disk is modeled using a hydrostatic pressure-based formulation by \citet{Blitz2006}, and for the spheroidal component star formation rates depend on the gas mass content and dynamical time of the system. Stellar feedback is implemented via a power-law relation between outflow rates and energy injection from the stellar population, with ejected gas stored in a reservoir and reincorporated on timescales proportional to the halo dynamical timescale. Summaries of these implementations are provided in \citet[Section 2.2]{2018MNRAS.474.5206K}, \citet[Section 2.1]{2023ApJ...948...87W}, and \citet[Appendix A]{2024MNRAS.529.3387A}. 

        With these physics included, the model naturally predicts that the occupation fraction of halos falls to nearly zero below a halo mass of about $3\times10^{7}~\Msol$, confirming that our merger trees have sufficient resolution for the regime relevant to this work. The model also reproduces key observables of Milky Way satellites, including the luminosity function, the size–mass relation, and the velocity dispersion–mass relation for classical and ultra-faint systems (see \citealt{2024MNRAS.529.3387A}) and we further examine its performance under the modified dark matter physics explored here (see Appendix~\ref{sec:distributions}).

        We consider WDM models with masses of 3, 6, 10, 20, and 40~keV, which set the matter power spectrum as described by \citet{2023PhRvD.108d3520V}. We also consider a reference CDM model. For each dark matter model, we use \textsc{Galacticus} to generate full subhalo populations for 10 Milky Way-like galaxies.

        \section{Analogues of Milky Way satellites}\label{sec:analogues}

        For each of the three galaxies Segue~1, Tri~II, and Tuc~V, and for each dark matter scenario, we randomly draw $10^4$ satellite galaxies from the 10 Milky Way-like galaxies modeled by \textsc{Galacticus}.\footnote{Our draws include roughly 500 unique Segue~1 analogues, 100 unique Tri~II analogues, and 2000 unique Tuc~V analogues, with some variation depending on the dark matter model.}
        These galaxies are drawn with replacement with draw probabilities proportional to the measurement distributions of the absolute magnitude $M_V$, the half-light radius $r_\mathrm{h}$, and the orbital pericenter radius $r_\mathrm{p}$ (from section~\ref{sec:satellites}).
        For each galaxy, this procedure corresponds to adopting a version of the measurement distribution where the \textsc{Galacticus} satellite distribution is taken as a prior.
        In appendix~\ref{sec:distributions}, we examine the degree to which the satellite galaxy distribution in \textsc{Galacticus} matches the properties of Segue~1, Tri~II, and Tuc~V. We find that \textsc{Galacticus} tends to produce galaxies with significantly lower $r_\mathrm{h}$, possibly because it does not currently apply tidal heating to the stellar distribution. However, the \textsc{Galacticus} satellite distribution is otherwise consistent with the properties of these galaxies.

\begin{figure}
			\centering
			\includegraphics[width=\columnwidth]{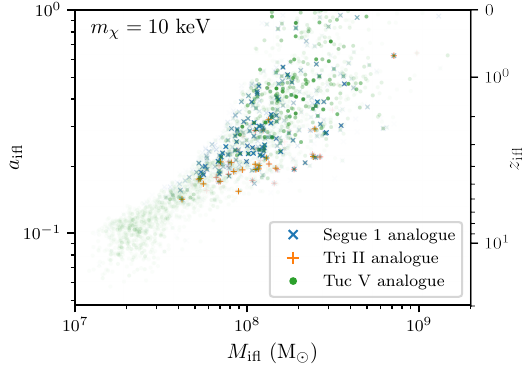}
			\caption{\textsc{Galacticus} analogues of Segue~1, Tri~II, and Tuc~V for 10~keV WDM. We show their distribution in terms of the halo virial mass $M_\mathrm{ifl}$ at the infall redshift $z_\mathrm{ifl}$ (or scale factor $a_\mathrm{ifl}$). Darker symbols represent galaxy analogues that are drawn more times and hence weighted more strongly in our analysis.}
			\label{fig:analogues_Mz}
		\end{figure}

        For 10~keV WDM, figure~\ref{fig:analogues_Mz} shows the halo masses and redshifts of the \textsc{Galacticus} analogues of Segue~1, Tri~II, and Tuc~V at their time of first infall into a host halo. We use a spherical-overdensity mass definition with the overdensity set by the spherical collapse model \citep[][]{1998ApJ...495...80B}. These galaxies span a broad range, but the distributions are centered at around a halo mass of $M_\mathrm{ifl}\sim 10^8~\Msol$ at an infall redshift of $z_\mathrm{ifl}\sim 3$.

        \subsection{Prompt cusps}

\begin{figure}
			\centering
			\includegraphics[width=\columnwidth]{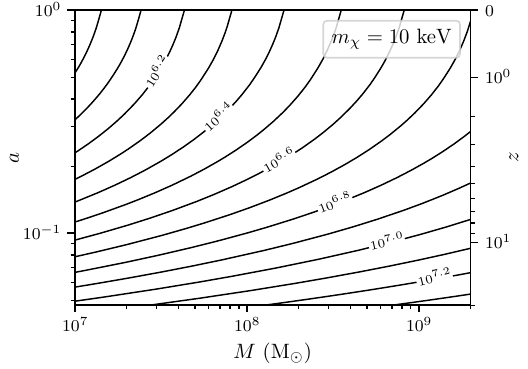}
			\caption{
            Cusp coefficient $A$ as a function of halo mass $M$ and scale factor $a$ (or redshift $z$), shown on the same axes as figure~\ref{fig:analogues_Mz}. The contours indicate the median $A$ in units of $\Msol\kpc^{-1.5}$ predicted by the cusp-halo relation of \citet{2025ApJ...993...93D} for WDM with $m_\chi=10$~keV.
            }
			\label{fig:A_Mz}
		\end{figure}

        We now turn to the prompt cusps at the centers of these galaxies' halos. Their coefficients $A$ are set by the cusp-halo relation of \citet{2025ApJ...993...93D}, which predicts $A$ as a function of halo mass, time, and cosmology. Figure~\ref{fig:A_Mz} visualizes this relation for 10~keV WDM on the same mass--redshift axes as figure~\ref{fig:analogues_Mz}, with $A$ rising toward higher halo mass and higher redshift. As described in section~\ref{sec:model}, \textsc{Galacticus} assigns each halo its cusp from this relation at the mass $M_\mathrm{res}=10^7~\Msol$ and redshift $z_\mathrm{res}$ at which the halo crosses the resolution limit, rather than at its infall mass and redshift.
        
\begin{figure}
			\centering
			\includegraphics[width=\columnwidth]{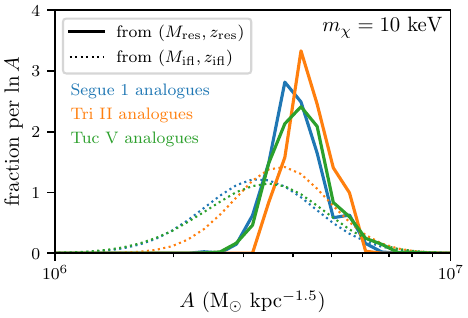}
			\caption{Distribution of the prompt cusp coefficients $A$ for \textsc{Galacticus} analogues of Segue~1, Tri~II, and Tuc~V in a 10~keV WDM scenario. The solid lines use the \textsc{Galacticus} prompt cusp prescription, which assigns the cusp at the mass $M_\mathrm{res}=10^7~\Msol$ and redshift $z_\mathrm{res}$ at which each halo crosses the resolution limit. These cusps are biased relative to those that would result from using the mass $M_\mathrm{ifl}$ and redshift $z_\mathrm{ifl}$ at subhalo infall, but we show in appendix~\ref{sec:simulations} that this bias is physically correct.}
			\label{fig:analogues_A}
		\end{figure}

        Figure~\ref{fig:analogues_A} shows the coefficients $A$ of the $\rho=Ar^{-1.5}$ prompt cusps of these galaxies as determined by \textsc{Galacticus} for 10~keV WDM.
        The cusps are relatively narrowly distributed about $A\sim 4\times 10^6~\Msol\kpc^{-1.5}$. For example, this means that the density at radii $r\sim 25~\pc$ is $\rho\sim 1~\Msol\pc^{-3}$. Note that $25~\pc$ is approximately the three-dimensional half-light radius of Segue~1, Tri~II, and Tuc~V.

        \begin{figure*}
			\centering
			\includegraphics[width=\textwidth]{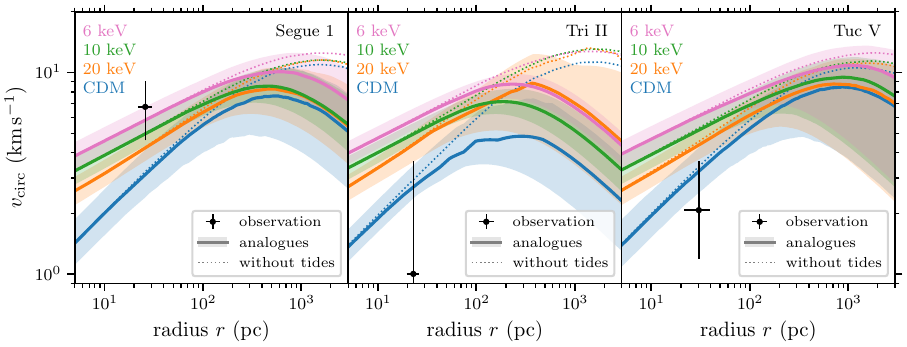}
			\caption{Radial profiles of the circular orbit velocity $v_\mathrm{circ}$ for \textsc{Galacticus} analogues of Segue~1 (left-hand panel), Tri~II (middle panel), and Tuc~V (right-hand panel). The solid curves show the median value at each radius, while the shading marks the $1\sigma$ (68 percent) scatter. Different colors correspond to different dark matter models. For comparison, the dotted curves show the median profiles prior to tidal evolution. The black points with error bars represent the $1\sigma$ range of the observationally inferred circular orbit velocity at the half-light radius, per section~\ref{sec:satellites}.}
			\label{fig:vc_profiles}
		\end{figure*}
        
        For comparison, the dotted curves in figure~\ref{fig:analogues_A} show the distributions of prompt cusps assigned using the cusp-halo relation at the mass $M_\mathrm{ifl}$ and redshift $z_\mathrm{ifl}$ at which each halo first falls into a host (i.e., comparison of figures \ref{fig:analogues_Mz} and~\ref{fig:A_Mz}).
        Conceptually, these can be interpreted as the cusp coefficients of field halos with similar properties to the halos of our Segue~1, Tri~II, and Tuc~V analogues.
        The distributions evidently differ to a significant degree.
        Cusp coefficients $A$ drawn from \textsc{Galacticus} (using $M_\mathrm{res}$ and $z_\mathrm{res}$) are higher by about 20 percent than those derived from $(M_\mathrm{ifl},z_\mathrm{ifl})$, and they are significantly less scattered.

        The higher $A$ of \textsc{Galacticus} subhalos is physically appropriate and reflects the distinction between newly infalling subhalos and field halos overall. Prior to infall into a host, a subhalo is simply a field halo in a dense environment. Halos in dense environments are known to be \textit{assembly biased} \citep[e.g.][]{2020MNRAS.493.4763M}, such that they tend to have grown more slowly and hence are ``older''. Such halos formed their prompt cusps earlier, so these cusps have higher density coefficients $A$. We show in appendix~\ref{sec:simulations} that halos which are about to fall into a much larger host indeed have cusp $A$ around 20 percent higher than is typical for field halos at the same mass and time.\footnote{However, assembly bias does not naturally emerge from the excursion set methods used by \textsc{Galacticus} \citep[nor does it emerge from excursion set theory with correlated steps, as we verified using the methods of][]{2024MNRAS.528.1372D}. Consequently, while \textsc{Galacticus} produces approximately the correct median $A$, it is not clear that this is due to physically correct modeling.} 

        On the other hand, the suppressed scatter of \textsc{Galacticus} subhalo cusp coefficients is not in accordance with simulation results.
        We show in appendix~\ref{sec:simulations} that newly infalling subhalos have similar scatter in $A$ to field halos of the same mass.
        Since the scatter in $A$ largely arises from scatter in halo mass accretion histories, this outcome suggests that \textsc{Galacticus} mass accretion histories may exhibit too little scatter.
        
        To correct for this effect, we add scatter to the \textsc{Galacticus} subhalo cusp coefficients $A$ by scaling each $A$ by a random variable $\delta_A$. We take $\delta_A$ to be log-normally distributed with 95 percent of the total (time-dependent) scatter in $A$ given in appendix~\ref{sec:simulations}. With this modification, the scatter in the \textsc{Galacticus} $A$ matches the scatter in the $A$ derived from $(M_\mathrm{ifl},z_\mathrm{ifl})$.

        \subsection{Circular velocity profiles}

        With the scatter correction applied, figure~\ref{fig:vc_profiles} shows the radial profiles of the circular orbit velocity for our Segue~1, Tri~II, and Tuc~V analogues in a range of dark matter scenarios.\footnote{For figure~\ref{fig:vcirc}, we consider the circular orbit velocity for Segue~1 analogues, Tri~II analogues, and Tuc~V analogues all together.} As noted above, we use the cusp-NFW density profile in equation~(\ref{cuspNFW}) modified by the tidal heating prescription of \citet{2014ApJ...792...24P} and \citet{2020MNRAS.498.3902Y}.
        The circular orbit velocity is 
        \begin{equation}
            v_\mathrm{circ}(r)
            =
            \sqrt{G M(r)/r},
        \end{equation}
        where $M(r)=\int_0^r 4\pi r'^2\rho(r')\diff r'$ is the mass enclosed within the radius $r$.
        For WDM, figure~\ref{fig:vc_profiles} shows how prompt cusps greatly boost orbital velocities near the center of the system due to how they concentrate more dark matter mass there. For lighter dark matter, the prompt cusps are denser, leading to higher $v_\mathrm{circ}$.
        
        For reference, we also show (dotted curves) the circular velocity profiles prior to tidal evolution.
        Analogues of Segue~1, Tri~II, and Tuc~V generally have comparable circular velocity profiles prior to tidal evolution. However, the low orbit of Tri~II means the Tri~II analogues tend to be severely tidally stripped, while Segue~1 and Tuc~V analogues are affected by tidal forces to a smaller degree.

        Finally, the points with error bars in figure~\ref{fig:vc_profiles} show the observationally inferred circular orbit velocity of each galaxy at the half-light radius, $v_\mathrm{circ}(r_\mathrm{h})$, as discussed in section~\ref{sec:satellites}.
        Apparently, Tri~II and Tuc~V have $v_\mathrm{circ}(r_\mathrm{h})$ consistent with CDM and unusually low for WDM scenarios.
        We also emphasize that even for a dark matter mass as high as 20~keV, there is a significant difference from CDM in the expected $v_\mathrm{circ}(r_\mathrm{h})$ for all three galaxies, even though the current measurement uncertainty may nevertheless be too high to discern the models.
        
        Meanwhile, Segue~1 has $v_\mathrm{circ}(r_\mathrm{h})$ consistent with WDM and unusually high for CDM. However, this comparison comes with a caveat. Just as subhalos tend to have higher $A$ than average, they also tend to be more concentrated. As we show in appendix~\ref{sec:simulations}, subhalos at the time of infall tend to have concentration parameters around 20-30 percent higher than the median of field halos of the same mass. However, \textsc{Galacticus} is calibrated to reproduce the concentration-mass relation of \citet{2016MNRAS.460.1214L} even for subhalos at the infall time. If the subhalo assembly bias were accounted for, we expect $v_\mathrm{circ}(r_\mathrm{h})$ for the CDM (and to a lesser extent WDM) analogues of Segue~1, Tri~II, and Tuc~V to be considerably higher.

        \section{Limits on warm dark matter}\label{sec:limits}

\begin{figure}
			\centering
			\includegraphics[width=\columnwidth]{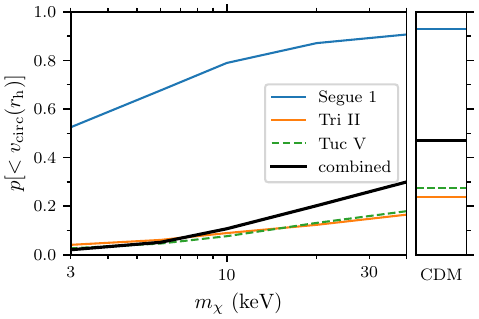}
			\caption{Probability of the circular orbit velocity at the half-light radius, $v_\mathrm{circ}(r_\mathrm{h})$, being at least as low as the measured value. We show this probability as a function of the WDM particle mass, with the the narrow panel on the right representing CDM. The three colored curves correspond to different dwarf galaxies, while the black curve shows the combined probability.}
			\label{fig:prob}
		\end{figure}

        For Segue~1, Tri~II, and Tuc~V, we now quantify how extreme the observationally inferred $v_\mathrm{circ}(r_\mathrm{h})$ are as a function of the dark matter scenario.
        For each galaxy and each dark matter scenario, using the error distribution of the observationally inferred $v_\mathrm{circ}(r_\mathrm{h})$ and the distribution of the \textsc{Galacticus} analogues, we evaluate the probability $p[<v_\mathrm{circ}(r_\mathrm{h})]$ of the circular orbit velocity lying below the measured value. Figure~\ref{fig:prob} shows these probabilities.
        For $m_\chi<10$~keV, the observationally inferred $v_\mathrm{circ}(r_\mathrm{h})$ of Tri~II and Tuc~V are each in the bottom 10th percentile with respect to the distribution of their \textsc{Galacticus} analogues. Meanwhile, Segue~1 is in the upper 10th percentile for CDM.
        
        We now evaluate the combined probability of all three galaxies having $v_\mathrm{circ}(r_\mathrm{h})$ at least as low as the measured values. To do this, we use
        \begin{equation}
            \chi^2 = -2\sum_{i}\ln p_i,
        \end{equation}
        where $p_i=p[<v_\mathrm{circ}(r_\mathrm{h})]$ for each galaxy $i$. The parameter $\chi^2$ is $\chi^2$-distributed with 6 degrees of freedom (for 3 galaxies), so we use the corresponding cumulative distribution function to obtain the desired combined probability. The black curve in figure~\ref{fig:prob} shows this combined probability.

        For $m_\chi\lesssim 5.8$~keV, the probability of all three $v_\mathrm{circ}(r_\mathrm{h})$ being as low as their measured values is less than 5 percent. For $m_\chi\lesssim 9.4$~keV, the probability is less than 10 percent. Hence, the kinematics of the faintest Milky Way satellites constrain $m_\chi>5.8$~keV at 95 percent confidence and $m_\chi>9.4$~keV at 90 percent confidence.
        These limits depend on the cusp coefficients $A$ that our modeling approach in section~\ref{sec:model} assigns to the observed galaxies, which could be refined with further model development. In appendix~\ref{sec:varyA}, we test how the limits depend on the assumed $A$.

		\section{Conclusion}\label{sec:conclusion}

        WDM models give rise to massive $\rho\propto r^{-1.5}$ prompt cusps at the centers of dark matter halos. The presence of these dark matter density cusps at the centers of galaxies would lead to higher orbital speeds relative to CDM scenarios. By analyzing the measured kinematics of the three smallest and faintest confirmed Milky Way satellite galaxies -- Segue~1, Tri~II, and Tuc~V -- we constrain the dark matter mass to be $m_\chi>5.8$~keV at 95 percent confidence or $m_\chi>9.4$~keV at 90 percent confidence.

        These results were enabled by the recent development of the \textit{cusp-halo relation} \citep{2025ApJ...993...93D} describing how to model the central prompt cusp of each dark matter halo.
        Based on this development, we implemented prompt cusps in the \textsc{Galacticus} semi-analytic model and used it to generate Milky Way satellite galaxy populations in a range of WDM scenarios.
        We then obtained constraints on WDM by comparing the observationally inferred properties of Segue~1, Tri~II, and Tuc~V to those of \textsc{Galacticus} analogues of these galaxies.

        Our 95-percent-confidence constraint is comparable to the strongest previous limits on WDM.
        However, we emphasize that the limits in this work are based on a fundamentally different principle.
        WDM is expected to suppress the abundance of galaxies, dark matter halos, or density perturbations on small scales, and previous works placed constraints on these suppressive effects. In contrast, the prompt cusps that we constrain represent an enhancement to small-scale structure that is expected to arise in WDM scenarios.
        The strength of our constraints with only three galaxies demonstrates the power of prompt cusps as a cosmological probe.

        We identified ways in which our modeling could be improved. The half-light radii $r_\mathrm{h}$ of \textsc{Galacticus} systems otherwise similar to Segue~1, Tri~II, and Tuc~V tend to be around half the observed $r_\mathrm{h}$. The match could be improved by properly including the influence of tidal heating on the stellar distribution. More thorough model calibration would also be beneficial \citep[e.g.][]{2025arXiv250900143R}. Since orbital velocities increase with radius, we expect that this correction would strengthen our limits on WDM, although the effect will be small since we weight our selection of \textsc{Galacticus} systems based on the observed $r_\mathrm{h}$.
        
        Other potential improvements concern \textsc{Galacticus}'s modeling of the dark matter halos hosting these satellites. We found that subhalos are assembly biased at their time of infall, such that they are internally denser than field halos of the same mass. While \textsc{Galacticus} produces subhalos with appropriately biased prompt cusps, it does not include the effect of assembly bias on halo concentrations. Accurately incorporating this effect \citep{2026arXiv260518940D} would raise orbital speeds, also strengthening our limits on WDM. Separately, \textsc{Galacticus}'s prediction for the halo hosting each observed satellite -- and especially that halo's pre-infall mass growth history, which sets the cusp coefficient $A$ -- could likewise be refined. We examine this dependence in appendix~\ref{sec:varyA}; improvements to the modeling of the galaxy--halo connection would help sharpen our limits.

        On the observational side, our constraints are mostly limited by measurement precision in the inferred kinematics of Segue~1, Tri~II, and Tuc~V \citep[for a discussion of sources of uncertainty, see][]{2019ARA&A..57..375S}. Significant differences from CDM are expected in these systems at observationally accessible radii even for dark matter masses $m_\chi\gtrsim 20$~keV. Consequently, limits on WDM could be greatly improved with more precise measurements of the kinematics of these systems. Limits could also be improved by including additional low-luminosity galaxies of comparable (or smaller) spatial sizes.

        An alternative direction is to search for prompt cusps in the resolved dark matter density profiles of dwarf galaxies \citep[e.g.][]{2023MNRAS.522L..78D}. Our analysis used only a single kinematic constraint, $\vc(r_\mathrm{h})$, which is sensitive to a prompt cusp's central mass contribution but not to its $\rho\propto r^{-1.5}$ shape. In dwarf spheroidal galaxies with richer kinematic data, the density profile shape could in principle be recovered with methods such as stellar dynamical modeling \citep[e.g.][]{2011ApJ...742...20W,2013ApJ...763...91J,2013A&A...558A..35B,2018MNRAS.481..860R,2019MNRAS.484.1401R,2020ApJ...904...45H,2024MNRAS.532.4157A,2024ApJ...970....1V}, simulation-based inference \citep{2023PhRvD.107d3015N,2025MNRAS.541.2707N}, and symbolic regression \citep{2026arXiv260105203M}. Such a search would, however, face two complications. First, the $\rho\propto r^{-1.5}$ signature is most distinctive well inside $r_\mathrm{h}$, where stellar tracers are sparse and inferences are fragile \citep[e.g.][]{2018MNRAS.474.1398G,2018ApJ...860...56S,2021MNRAS.507.4715C}. Second, dwarfs with the richest kinematic data have many more stars than those we studied, so baryonic feedback may have modified their inner dark matter distributions \citep{2019MNRAS.484.1401R}, complicating the comparison to the unperturbed cusp expectation.
		
		\begin{acknowledgments}
			MSD thanks Stacy Kim and Nondh Panithanpaisal for helpful discussions. Computations for this work were carried out on the OBS HPC computing cluster at the Observatories of the Carnegie Institution for Science.
		\end{acknowledgments}

		\appendix

        \section{Halos and cusps in simulations}\label{sec:simulations}

        Here we analyze the simulations of \citet{2023MNRAS.518.3509D} using the methodology of \citet{2025ApJ...993...93D} to obtain some conclusions about the scatter and bias of subhalo density profiles. There are three simulations with different initial matter power spectra labeled as $n=-2.67$, $n=-2$, and $n=1$. Following \citet{2025ApJ...993...93D}, we parametrize the cosmic time in terms of the rms linear-theory density contrast, $\sigma_0$, and we use the natural mass unit $\bar\rho(\sigma_0/\sigma_2)^{3/2}$. Here $\bar\rho$ is the average matter density and
        \begin{equation}
        \sigma_j^2 = \int \frac{\diff^3\vec{k}}{(2\pi)^3}P(k) k^{2j},
        \end{equation}
        where $P(k)$ is the linear-theory matter power spectrum.
        For reference, for $m_\chi=(6,10,20)$~keV, $\bar\rho(\sigma_0/\sigma_2)^{3/2}\simeq (1.1\times 10^6,1.9\times 10^5,2.0\times 10^4)~\Msol$ and $\sigma_0\simeq (8.6,9.8,11.3)a$ during matter domination, where $a$ is the scale factor (normalized to $a=1$ today).

\begin{figure}
			\centering
			\includegraphics[width=\columnwidth]{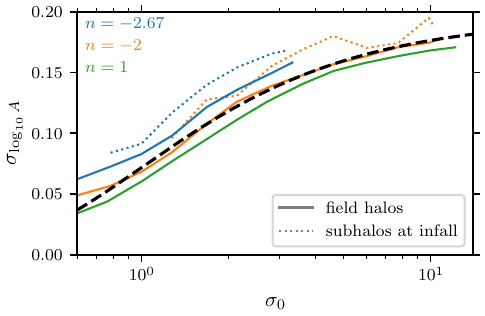}
			\caption{68 percent scatter in cusp coefficients $A$ at fixed halo mass, shown as a function of time. The solid curves include all field halos, while the dotted curves are restricted to halos that will fall into another halo at least 100 times more massive in the next 3-6 percent of a Hubble time. The dashed curve represents equation~(\ref{scatter}).}
			\label{fig:scatter_A}
		\end{figure}

        \paragraph{Scatter in $A$ at early times}
		\citet{2025ApJ...993...93D} noted that at sufficiently late times, there is about 0.15 dex scatter in cusp coefficients $A$ of field halos at fixed halo mass. However, the scatter is lower at early times. Figure~\ref{fig:scatter_A} shows the 68 percent scatter in $A$ for mass bins in which the median $A$ is larger than $\tilde A|_{\sigma_0=1}$, the typical coefficient for cusps forming at the time when $\sigma_0=1$.\footnote{While the scatter depends only weakly on halo mass for $A> \tilde A|_{\sigma_0=1}$, it becomes significantly mass-dependent when $A\lesssim \tilde A|_{\sigma_0=1}$; see \citet{2025ApJ...993...93D}. However, such low-$A$ cusps typically correspond to halo masses that are too low to be relevant for galaxies.} As a function of the time parameter $\sigma_0$, it is approximated well by
		\begin{align}\label{scatter}
			\sigma_{\log_{10}A} = 0.195\e^{-1/\sigma_0},
		\end{align}
		or equivalently $\sigma_{\ln A} = 0.45\e^{-1/\sigma_0}$.
		
\begin{figure}
			\centering
			\includegraphics[width=\columnwidth]{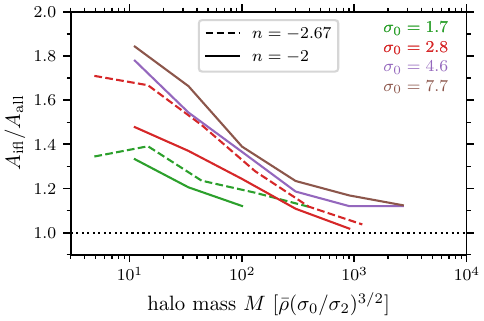}
			\caption{Bias in cusp coefficient $A$ for halos that will fall into another halo at least 100 times more massive in the next 3-6 percent of a Hubble time. As a function of halo mass and time (parametrized by $\sigma_0$; different colors), we show the ratio between the median $A$ for these halos and the median $A$ for all field halos. We restrict our consideration to halos of at least 32 simulation particles, we use halo mass bins of width $\Delta\ln M=1.1$, and we only include mass bins with at least 30 halos.}
			\label{fig:bias_A}
		\end{figure}

		\paragraph{Bias of subhalo cusps}
        We now consider field halos that will first merge with another halo 2 simulation snapshots in the future \citep[according to the friends-of-friends halo definition used in][]{2025ApJ...993...93D}. Snapshots are spaced by about 3 percent in the scale factor, so we consider halos that merge in the next 3--6 percent of a Hubble time. We additionally restrict our consideration to mergers for which the other halo is at least 100 times as massive as the halo under consideration. We first analyze the cusp coefficients $A$ of these soon-to-infall halos. Figure~\ref{fig:bias_A} compares the median $A$ for these halos to the median $A$ for all field halos in the same mass bin. Evidently, halos that are about to merge onto a larger system have considerably higher $A$ than is typical. Meanwhile, the dotted curves in figure~\ref{fig:scatter_A} show the scatter in $A$ for this sample; it is evidently comparable to the scatter in $A$ for field halos overall.

\begin{figure}
			\centering
			\includegraphics[width=\columnwidth]{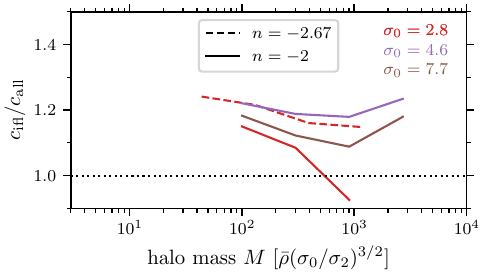}
			\caption{Similar to figure~\ref{fig:bias_A} but showing the bias in halo concentration $c$ instead. Here, due to the need to resolve internal structures, we restrict our consideration to halos of at least 300 simulation particles.}
			\label{fig:bias_c}
		\end{figure}

		\paragraph{Bias of subhalo concentrations}
        For the same sample of halos that are about to merge onto larger systems, figure~\ref{fig:bias_c} compares their median concentration parameter $c$ to the median $c$ for all field halos of the same mass. Evidently, halos that are about to merge onto a larger system also have considerably higher $c$ than is typical. As in \citet{2025ApJ...993...93D}, we define $c=r_{200}/r_{-2}$, where $r_{200}$ is the radius enclosing 200 times the average matter density (centered on the minimum of the gravitational potential) and we estimate $r_{-2}\simeq r_\mathrm{max}/2.2$, where $r_\mathrm{max}$ is the radius of maximum circular orbit velocity.

        \section{Further observable properties of satellite galaxy analogues}\label{sec:distributions}
        
        \begin{figure*}
			\centering
			\includegraphics[width=\textwidth]{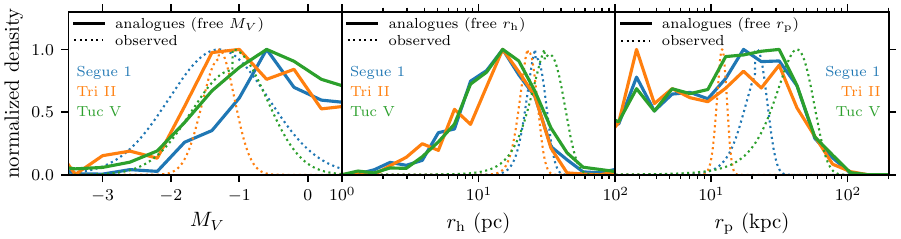}
			\caption{Absolute $V$-band magnitude $M_V$, half-light radius $r_\mathrm{h}$, and pericenter radius $r_\mathrm{p}$ for \textsc{Galacticus} analogues of Segue~1, Tri~II, and Tuc~V. The left-hand panel shows the distribution of $M_V$ for analogues selected by $r_\mathrm{h}$ and $r_\mathrm{p}$; the middle panel shows the distribution of $r_\mathrm{h}$ for analogues selected by $M_V$ and $r_\mathrm{p}$; and the right-hand panel shows the distribution of $r_\mathrm{p}$ for analogues selected by $M_V$ and $r_\mathrm{h}$. For comparison, the dotted curves show the uncertainty distributions of the observationally inferred values (which can exhibit minor discontinuity due to how we model asymmetric uncertainties).}
			\label{fig:anadist}
		\end{figure*}

        Here we compare the distribution of \textsc{Galacticus} analogues of Segue~1, Tri~II, and Tuc~V to the observed properties of these systems.
        Similarly to in section~\ref{sec:analogues}, we sample analogues of these galaxies by weighting according to the uncertainty distributions of the observationally inferred parameters $M_V$, $r_\mathrm{h}$, and $r_\mathrm{p}$. However, we make one change: we weight by the distributions of only two of the three parameters, leaving the third parameter free.
        This procedure allows us to test how typical Segue~1, Tri~II, and Tuc~V are with respect to galaxies in \textsc{Galacticus}.

        For 10~keV WDM, figure~\ref{fig:anadist} shows the separate distributions of $M_V$, $r_\mathrm{h}$, and $r_\mathrm{p}$ obtained for Segue~1, Tri~II, and Tuc~V analogues through this procedure. Several noteworthy differences are visible between these galaxies and the \textsc{Galacticus} analogues.
        
        \paragraph{Absolute magnitudes $M_V$}
            \textsc{Galacticus} predicts a large population of galaxies similar to Segue~1, Tri~II, and Tuc~V but with much lower luminosity (higher $M_V$). This outcome is naturally explained as an observational bias: fainter galaxies are more difficult to find and identify. See \citet{2025arXiv251115808A} for further discussion of ``hyper-faint'' galaxies in \textsc{Galacticus}.
        \paragraph{Half-light radii $r_\mathrm{h}$}
            \textsc{Galacticus} galaxies similar to Segue~1, Tri~II, and Tuc~V tend to be spatially smaller. Observationally, there are numerous systems of comparable $M_V$ and significantly lower $r_\mathrm{h}$ ($\lesssim 10$~pc), but whether they are galaxies or star clusters is uncertain; \textsc{Galacticus} could be predicting that many of these systems are galaxies \citep[see also][]{2025arXiv251115808A}. On the other hand, \textsc{Galacticus} could be systematically underestimating $r_\mathrm{h}$, possibly because it currently applies tidal heating only to dark matter and not to stars.
        \paragraph{Pericenter radii $r_\mathrm{p}$}
            Segue~1, Tri~II, and Tuc~V span a range of $r_\mathrm{p}$ consistent with the broad distribution predicted by \textsc{Galacticus}. However, \textsc{Galacticus} also predicts the presence of a numerous similar galaxies with $r_\mathrm{p}\ll 10$~kpc. This outcome is likely artificial. Galaxies with such low pericenters should be severely tidally stripped, suppressing their luminosity (and sometimes disrupting them entirely), but \textsc{Galacticus} currently only applies tidal stripping to dark matter and not stars.

        \section{Sensitivity to the cusp coefficient $A$}\label{sec:varyA}

\begin{figure}
			\centering
			\includegraphics[width=\columnwidth]{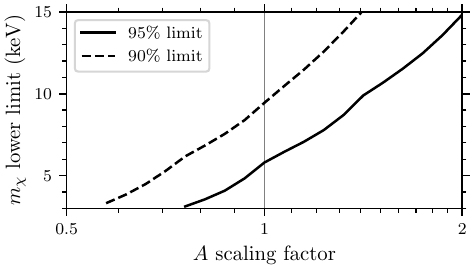}
			\caption{
            Lower limits on the WDM particle mass $m_\chi$ when the prompt cusp coefficients $A$ of all \textsc{Galacticus} analogues are multiplied by a common scaling factor. The solid and dashed curves show the 95 and 90 percent confidence limits, respectively. A scaling factor of unity (vertical line) corresponds to the fiducial analysis of section~\ref{sec:limits}.
            }
			\label{fig:A_scale}
		\end{figure}

        To assess how strongly our results depend on the modeling of the prompt cusps, we repeat the analysis of section~\ref{sec:limits} with the cusp coefficient $A$ of every analogue multiplied by a common scaling factor. Figure~\ref{fig:A_scale} shows the resulting 95 and 90 percent lower limits on the WDM particle mass $m_\chi$ as a function of the scaling factor. A scaling factor of unity recovers our fiducial limits of 5.8 and 9.4~keV, and the limits depend strongly on $A$: a 30 percent reduction weakens the 95 percent limit to around 3~keV, while a 30 percent increase strengthens it to over 8~keV.

        An ingredient of our modeling with a significant potential to introduce a systematic offset in $A$ is the galaxy--halo connection. Alternative galaxy--halo connections for ultra-faint dwarfs include the empirical $M_V$--$M_\mathrm{peak}$ relation of \citet{2020ApJ...893...48N} and the GRUMPY semi-analytic model of \citet{2022MNRAS.516.3944M} \citep[see also][]{2025arXiv251121824B}. \citet{2024MNRAS.529.3387A} find that \textsc{Galacticus} places the faintest observed satellites in halos a factor of $\sim 2$ heavier at peak mass than either alternative, attributed to its omission of pre-infall mass loss. At typical infall masses and times of Segue 1, Tri II, and Tuc V analogues (figure~\ref{fig:analogues_Mz}), the cusp-halo relation (figure~\ref{fig:A_Mz}) indicates that a factor-of-two change in halo mass at fixed epoch shifts $A$ by roughly 30 percent. Less massive halos host weaker cusps and would weaken our limits.

        Two considerations, however, complicate this estimate. First, the factor of two is in peak mass. In simulations, a halo begins to lose mass well outside its eventual host \citep{2014ApJ...787..156B}, reaching its peak mass before infall; \textsc{Galacticus}, lacking this effect, also dates the peak somewhat too late. Figure~\ref{fig:A_Mz} shows that a later epoch lowers $A$, opposing the increase from the overestimated mass. Second, we assign $A$ not at the peak but at the much earlier time the halo's main progenitor crosses the resolution limit (see section~\ref{sec:model}). Pre-infall mass loss acts late and changes neither this time nor which halos form galaxies, so to the extent the offset reflects it, $A$ is unaffected; the offset would bias $A$ only insofar as it instead reflects an error in the halos assigned to these galaxies. The net effect on $A$ is therefore unclear, and determining it will require further study of the galaxy--halo connection.

        \clearpage
		\bibliography{main}{}
		\bibliographystyle{aasjournalv7}
		
		
		
	\end{document}